\documentstyle[12pt]{article}

\def\ben{\begin{enumerate}}  \def\een{\end{enumerate}}
\def\beq{\begin{equation}}   \def\eeq{\end{equation}}
\def\bea{\begin{eqnarray}}  \def\eea{\end{eqnarray}}

\def\lsim{\raise0.3ex\hbox{$<$\kern-0.75em\raise-1.1ex\hbox{$\sim$}}}
\def\gsim{\raise0.3ex\hbox{$>$\kern-0.75em\raise-1.1ex\hbox{$\sim$}}}

\begin{document}

\vbox to 1 truecm {}
\begin{center}
{\large \bf A note on Measurement}
\par

\vspace{1 truecm}

{\bf Bernard d'Espagnat}\\
{\it Laboratoire de Physique Th\'eorique}\footnote{Unit\'e Mixte de Recherche
UMR 8627 - CNRS - Fax : (33) 1 69 15 82 87 }\\    {\it Universit\'e 
de Paris XI,
B\^atiment 210, 91405 Orsay Cedex, France}
\end{center}

\vspace{2 truecm}
\begin{abstract}
Grounded on the quantum measurement riddle, a general argument 
against the universal
validity of the superposition principle was recently put forward by 
Bassi and Ghirardi
\cite{1r}. It is pointed out that this argument is valid only within 
the realm of the
philosophy of ``objectivistic realism'' which is not a necessary part 
of the foundations
of physics, and that recent developments including decoherence theory 
do account for the
{\it appearance} of macroscopic objects without resorting to a break 
of the principle.
  \end{abstract}

\newpage
\pagestyle{plain}

How do we know that there is a stone on the path, or a tree in the 
courtyard? Obviously
(as many philosophers have kept stressing) by having a look. So that, 
if we were
extremely cautious not to make unwarranted statements we should not 
bluntly say that
there is a stone on the path (or a tree in the courtyard). We should 
say~: ``We know that
if we had a look at the path, to check whether or not we have the 
impression of seeing a
stone, we should actually get the impression in question''. As long 
as we remain within
the realm of pure thinking, this remark does not amount to taking an 
option for or
against objectivistic realism. It is just a matter of cautiousness, 
that is, of taking
care not to make unjustifiable claims. It may be that objectivistic 
realism is true.
But, since it is unprovable, it may also be that it is not. So, we 
keep on the safe side
by not implicitly postulating it. \par

In ordinary life making use of such long, intricate sentences is 
quite impossible. For
all practical purposes we are therefore fully justified - even if we 
are not diehard
realists - in using the shorter, so called ``realistic'', sentences, 
that describe
objects as ``really being'' here or there. In the quantum mechanical 
realm the situation,
however, is different. As everybody knows, this is a domain in which 
too ``realistic''
sentences, implicitly postulating that all the quantities of interest 
always {\it have}
values, would lead us astray. And we may well suspect that, when we 
assume quantum
mechanics is universal and apply it to macroscopic systems, something 
similar may be
true also concerning some sentences bearing on such systems. But 
still~: even  in the
realm of atomic and subatomic physics there is at least one 
circumstance in which the
use of ``realistic'' sentences - involving the verbs ``to have'' and 
``to be'' - is both
harmless and convenient. This is when {\it we} know (for sure) 
beforehand that, if we
measured an observable $B$ on a system $S$, we would get eigenvalue 
$b_k$ of $B$ as an
outcome. In that case we may assert that system $S$ is in a state 
described by one of the
eigenvectors of $B$ corresponding to eigenvalue $b_k$ (when $S$ is 
entangled with other
systems what we may assert is, more precisely, that the {\it overall 
composite system} is
in an eigenstate of $Bf1$ corresponding to eigenvalue $b_k$, where 
$Bf1$ stands for the
extension of operator $B$ to the Hilbert space of the whole composite 
system). \par

Similarly, when, as with the example of the stone on the path, we 
know for sure that if
we looked we would have the impression of seeing a certain physical 
system lying within
a given region of space instead of outside it, we are allowed to 
consider this knowledge
as enabling us to make some definite statements concerning the 
quantum mechanical
description of this system. For instance, when the system in question 
is an electron we
are allowed to infer from such a  knowledge that the state vector of 
the electron (or,
better to say, of the whole Universe including the electron) is an 
element of a certain
set of vectors. Or, to take another example, consider the well-known 
explicit model of a
measurement process that was given by von Neumann and which just 
consists of a ``spin''
and a one degree of freedom apparatus (see \cite{2r}, Chapter VI, ¤3 
or \cite{3r},
Section 14.3). When the initial spin state is given and is an 
eigenstate $u_k$ of the
quantity $S_z$ that the instrument has been instructed to measure, we 
can derive from the
Schr\"odinger equation the certainty that if, after the interaction, 
we had a look at the
position $B$ of the pointer we would get the ``outcome'' $b_k$. Then, 
according to the
above, we are allowed to infer from this knowledge a definite 
statement concerning the
pointer final state, namely that it is the state $f_k$ corresponding 
to $b_k$. However a
problem arises when we assume that initially, the spin state is {\it 
not} an eigenstate
of $S_z$. This problem - a conceptual one! - bears on the question 
whether or not we
should still assert that the pointer lies at some definite place, 
corresponding to one
of the $b_k$, and that its quantum state is therefore an element of 
one or other of the
corresponding definite sets of vectors. \par

More precisely, the problem consists in the fact that the above 
described general
approach can be applied in two different ways, depending on the 
amount of initial
knowledge we consider we have, and that these two ways lead to different
conclusions.\par

One of these ``ways of arguing'' - call it ``option A'' - consists in 
clinging to the
realist philosophy, which claims that the reason why we see 
macroscopic objects as
having definite forms and definite localizations in space is that 
they really exist as
such, quite independently of us, that is, of our sensorial and 
intellectual equipment.
We see them at definite places because they {\it are} at definite 
places. Hence we know
beforehand that a pointer, say, cannot be at the same time in two 
macroscopically
different states and of course this - very general but, nevertheless, 
quite certain -
knowledge must be taken into account when we apply the above 
described procedure in
order to determine the final state of the pointer or, better to say, 
in order to state
some general conditions that the state in question must fulfill. \par

The other ``way of arguing'' - call it ``option B'' - consists in 
keeping close to a
standpoint taken by a number of ancient Greek philosophers (Plato 
foremost), adopted as
a ``starting point'' (though finally dropped) by Descartes and 
forcefully argued for by
Kant. Roughly, this is the view that, quite generally, the 
testimonies of our senses are
deceitful and should not be taken at face value. More precisely, it 
consists in claiming
(contrary to Galileo, Descartes and Locke) that, when all is said and done, the
``qualities'' Locke called ``primary'' (shape, position, motion, 
etc.) should be
considered as being man-dependent precisely in the same sense as are 
those he called
``secundary'' (colour, taste, smell, etc.: the taste of a fruit 
depends on the fruit but
it also depends on us). According to this trend of thought 
(considered as being the most
reasonable one by, perhaps, the majority of contemporary 
philosophers), the fact that we
perceive such ``things'' as macroscopic objects lying at distinct 
places is due, partly
at least, to the structure of our sensory and intellectual equipment. 
We should not,
therefore, take it as being part of the body of sure knowledge that 
we have to take into
account for defining a quantum state. \par

The branch of study conventionally called ``measurement theory'' 
almost entirely
developed within option A and was an attempt at showing that the said option is
compatible with conventional quantum mechanics and the completeness 
assumption. In their
recent article [1] Bassi and Ghirardi referred to a book \cite{3r} in 
which I reviewed
and discussed the main proposals that were put forward to that end 
and pinpointed the
considerable difficulties they all must cope with. The main one of 
these is of course
that when the system $S$ on which a quantity $B$ is to be measured is 
not, initially, in
an eigenstate of $B$, if a state vector is initially attributed to 
the pointer the
Schr\"odinger time evolution leads, for the overall system $S$ 
composed of $S$ and the
pointer (or of $S$ and the rest of the world if, along with the 
pointer, we take the
environment into account, as we should), to a state that is a superposition of
macroscopically distinct states; a result which is incompatible with 
option A as noted
above. In order to overcome this difficuly it was stressed in 
\cite{3r}, in particular,
that initially describing the pointer (or the pointer-environment 
system) by means of a
state vector is a considerable idealization, and that, because of our 
ignorance of its
detailed atomic structure and so on, it should actually be 
represented by a density
matrix, with the consequence that the final state of $S$ would also 
be represented by
such a matrix. In view of the fact that, in general, a given density 
matrix corresponds
not to one but to infinitely many proper mixtures it could then a 
priori be hoped that,
among the latter, some would be composed of states fulfilling the 
condition of not being
superpositions of macroscopically distinct states. The specific 
difficulty mentioned
above would then be removed (even if other ones conceivably 
remained). Also - let this
be added here - it could be hoped that, somehow, the apparent 
violation of determinism
characterizing such measurement processes could be reconciled with 
the deterministic
nature of the Schr\"odinger time evolution by invoking the ignorance 
probabilities
inherent in proper mixtures. However, as pointed out in \cite{3r}, 
the result of the
investigations in question was that the first of these two hopes is, 
in fact, unfounded,
in the sense that, concerning the final state of $S$, in the 
considered situation proper
mixtures with the requested properties do not exist. Now, Bassi and 
Ghirardi gave a new
proof of this, and it may be considered that theirs is both simpler 
and more general. It
is true that they did not explicitely consider proper mixtures but it 
could be argued
that their proof applies separately to every component of such 
mixtures. Also, they did
not explicitely consider the question of determinismÉ but after all, 
neither did I.
\par

From their result Bassi and Ghirardi inferred that ``to have a 
consistent picture one
must accept that in a way or another the linear nature of the 
dynamics must be broken''.
Now, is this conclusion inescapable and general (at least within a 
genuinely quantum
description, with the completeness assumption made)~? This is the 
question that must now
be addressed to. \par

Within Option A the conclusion seems inescapable indeed. But on the 
other hand none of
the schemes that materialize the break is as yet considered, for 
various reasons, as
being fully convincing. In particular, the one that was developed by 
Ghirardi et al.
\cite{4r} offers no other motivation for the proposed modification 
than the one just
explained above, namely the ``necessity'' we think we are in of 
describing macroscopic
systems as never being in superpositions of macroscopically different 
states. But, as
the forgoing already indicates, this is merely a philosophical 
requirement. In fact,
scientists most righly claim that the purpose of science is to describe human
experience, not to describe ``what really is''; and as long as we 
only want to describe
human experience, that is, as long as we are content with being able 
to predict what
will be observed in all possible circumstances, it must be granded 
that Option B is
enough. We need {\it not} postulate the existence - in some absolute sense - of
unobserved (i.e. not yet observed) objects lying at definite places in ordinary
3-dimensional space. Consequently we have no need for such a break in 
the linear nature
of the dynamics as the one Bassi and Ghirardi suggest. To introduce 
such a momentous
change in the scientific description merely on the basis of a 
philosophical conception
of our relationship with the World is a procedure that may be 
considered as far removed
from normal scientific practice. \par

In view of all this (combined with such experiments as the one of the 
Haroche group
\cite{5r}, which plead convincingly in favor of the universality of the quantum
mechanical predictive rules), taking up Option B seems more advisable 
than taking up
Option A. Within the realm of Option B it is, to repeat, not at all 
considered as
certain that, independently of ourselves, macroscopic objects exist 
``out there'' as we
see them (with precise locations and so on). What is considered as 
certain (or at least
as ``well established'') is a set of predictive rules that enable us 
to foresee what we
shall observe. This set involves, in particular, the rules - call 
them ``the M rules'' -
that apply within the so-called macroscopic domain and it so happens 
that {\it these}
rules can be translated in the language of the descriptive laws of 
classical physics,
that is in terms of statements interpretable as bearing on objects 
existing ``in the
outside World''. It is then, to repeat, most convenient to make use 
of such a language,
but we should not infer from this that the language in question 
necessarily describes
elements of anything that could be referred to as ``man-independent 
reality''. Perhaps it
does but perhaps it does not. Now, in this field the most significant 
recent development
consists in the fact that, due to the (universally existing) 
interaction between a
macroscopic system and its environment (including its ``internal'' 
one, that is, the set
of its atomic variables), it could be shown (i) that the (predictive) 
M rules follow
from the (predictive) basic quantum rules (see \cite{6r}, Chapters 6 
and 7) and (ii)
that, for macroscopic systems, the {\it appearances} are those of a 
classical world (no
interferences etc.), even in circumstances, such as those occurring in quantum
measurements, where quantum effects take place and quantum 
probabilities intervene (see
e.g. \cite{7r}). This is the true significance of decoherence theory. 
In other words,
this theory has no meaning within objectivistic realism and should 
not therefore be
understood as signifying that a ``real'' collapse occurs, when 
``real'' is understood in
the sense it has within the philosophy in question. But it remains 
true that decoherence
explains the just mentioned {\it appearances} and this is a most 
important result. It may
be considered as implying that, from a quite strictly scientific 
viewpoint\footnote{I
have here in mind a viewpoint that would be totally faithful to the (perhaps
unreachable~!) scientific ideal of keeping to what seems 
unquestionable within collective
human experience, namely the impressions we share, without any 
admixture of presupposed
ideas concerning the actual existence of the forms thus perceived 
(see Appendix).}, the
above mentioned Bassi and Ghirardi claim is not justified. As long as 
we remain within
the realm of mere predictions concerning what we shall observe (i.e. 
what will {\it
appear} to us) - and refrain from stating anything concerning 
``things as they {\it
must} be before we observe them'' - no break in the linearity of 
quantum dynamics is
necessary. \par

On the other hand, this conclusion should not be interpreted as meaning that
investigations bearing on the so called ``measurement theory'' have 
proven nothing. What
they proved (within the realm of the completeness assumption) is that 
we must {\it
either} accept the break {\it or} grant that man-independent reality 
- to the extent that
this concept is meaningful - is something more ``remote from anything 
ordinary human
experience has access to'' than most scientists were up to now 
prepared to believe
(although science formerly contributed decisively to making plausible 
the idea that
Reality is not at all what it looks like). This is an important result, to the
derivation of which the Bassi and Ghirardi paper unquestionably brought a very
significant contribution.

\newpage

\section*{Appendix}
\hspace*{\parindent}

	Now, in thus comparing options A and B, was I unfair to the 
former~? After reading a
preliminary version of this article Prof. G. C. Ghirardi reminded me 
that also option B
has its limitations, an important one proceeding from the fact that a 
(nonpure) density
matrix corresponds to several proper mixtures. Consequently (as 
pointed out by Joos
\cite{8r} and rediscovered independently by myself \cite{9r}) when, 
for example,
decoherence is applied to the localization of macroscopic objects 
(dust grains, say) it
does not suffice, by itself, to prove that in an ensemble of such 
objects each element
occupies - or will be seen as occupying -  some definite place. In 
other words, the
localization process is not just a consequence of the formalism. It 
is also due to our
human way of perceiving so that, if we stick to the conventional 
notion of ``states''
(states of ``systems'' or of ``the World'') we have to grant that 
within option B
perceptions are linked in quite a loose way with the said states (as 
described by
density matrices). As Prof. Ghirardi stressed to me, there is, after all, no
considerable difference between such a state of affairs and the 
ancient view of London
and Bauer and Wigner, according to which the wave function is reduced 
by an individual
conscious act of perception. \par

There is, I must admit, substantial truth in this remark. On the 
other hand, I claim
that this disadvantage of the decoherence approach is, if not 
completely removed, at
least considerably alleviated if option B is understood as centered 
on predictivity, as
sketched in the first paragraph of the present article. More 
precisely, although,
personally, I tend to view physics without metaphysics as being 
conceptually incomplete,
I consider nevertheless that we should be careful not to include some 
admixture of the
latter in the technicalities of the former. In particular (as already 
stressed in
\cite{3r}), physicists should be cautious when using the notion of 
``state'', which,
because of its role in ordinary language, has questionable 
metaphysical implications. I
observe that if, as physicists, we only worry about predicting what 
are our chances of
observing this or that, and if, correlatively, we impart to the word 
``state'' no other
meaning than that of designating a mathematical tool allowing for 
such predictions, we
meet with no ambiguities whatsoever. In particular, I am happy to 
know (from decoherence
theory and so on) that, for the said purpose, I can, without running into
inconsistencies, allegorically use in my daily life the descriptive language of
classical physics and commonsense. I claim that many of our 
(conceptual) worries in
quantum mechanics proceed from the fact that, since the descriptive 
language is simpler
and more congenial than any conceivable predictive one, we use it 
whenever possible and
then interpret it metaphysically (and erroneously), as describing 
``what really is''.
\par

	Now, some philosophers would carry such a ``radicalism'' to 
its bitter end. They tend to
reject any notion of a reality not identified with the set of our 
impressions and
predictions. Clearly, we should not go that far. And I grant that 
option B does have
one  inconvenience at least~: that of suggesting going over from Platonicism to
neo-Kantianism. Existence is prior to anything else, reference, 
prediction etc. But
existence as Being, not the existence of objects. The trouble is 
that, while, through
physics, Being informs us quite definitely of what it is not (e.g. it 
is not composed of
localized elements), it seems reluctant at letting us know what it 
truly is. \par

\end{document}